# Study of electronic and optical properties of $Si_{1-x}Ge_xO_2$ with i phase structure to find high-k

M. Tirandari [1]

[1]Department of Physics, Sharif University of Technology, Tehran 11155-9161, Iran

**Abstract**

In this work, the electronic and optical properties of $Si_{1-x}Ge_xO_2$ at equilibrium and non-equilibrium condition were calculated using a full potential linear augmented plane wave plus local orbital method. The effect of Si and Ge on electronic and optical properties of $Si_{1-x}Ge_xO_2$ is investigated. It was found that $Si_{1-x}Ge_xO_2$ compounds could be used as a high-K in electronics devices.

## 1. Introduction

High dielectric constant materials are expected to replace conventional $SiO_2$ gate oxide to continue the scaling down of silicon based semiconductor devices. Using them for gate oxide allows us to make tunnel oxide with reducing leakage current. Although the amorphous oxide does not have grain boundaries and is isotropic, polycrystalline oxide with more abrupt interfaces allows us to scale down [1].

There are numerous oxides with extremely dielectric constants or "high K", but suitable candidates have not been identified. Ouang and Ching have reported a new pure phase of $SiO_2$ in inverse $Ag_2O$ structure, which is called i phase. This new phase has an unusually high dielectric constant [2]. Sevick and Bulutay used i phase for $GeO_2$ and $SnO_2$ and their ternary alloys with $SiO_2$ [3]. They suggests that *i* phase of $Si_{0.5}Ge_{0.5}O_2$ has a higher dielectric constant and a wide band gape. Moreover, it is lattice matched to Si(100) face.

In this work, we investigated the electronic and optical properties of $Si_{1-x}Ge_xO_2$ alloys to find the other high-k and to see the effect of Si and Ge concentration on their properties.

## 2. Method and details of calculations

The self-consistent calculations were carried out using WIEN2k [4], a full potential linear augmented plane wave plus local orbital (FPL/APW+lo) method. Exchange correlations

---

Email: mehraneh.tirandari@gmail.com

are included using the generalized gradient approximation (GGA) of Perdew-Burke-Ernzerhof (PBE). The LAPW+lo method expands the Kohn-Sham orbital in atomic-like orbitals inside the muffin-tin (MT) spheres and plane waves in the interstitial region. We chose MT radius of oxygen 1.05 Å and for Si and Ge, 1.92 Å to 1.89 Å. Total energy as a function of k points was calculated in the irreducible Brillouin zone (IBZ) to ensure the accuracy of the calculations. It is shown that 27 and 64 k points are sufficient. Self-consistent calculation was considered to be converged when the total energy was 0.1 mRy. The BZ integration was carried out using the modified tetrahedron method [5].

The supercell approach was adopted to model the $Si_{1-x}Ge_xO_2$ alloys. A 12 atoms $Si_{4-n}Ge_nO_8$ supercell was used. The supercell structure is shown in Fig. 1. We have $SiO_2$ for n=0 which is *i* phase and its properties has been calculated by other researchers [2, 3]. It has a simple cubic structure with $Pn\bar{3}m$ space group. We have $Si_{0.75}Ge_{0.25}O_2$, $Si_{0.5}Ge_{0.5}O_2$ and $Si_{0.25}Ge_{0.75}O_2$ for n=1, 2, 3 respectively. Different atomic sites for Si and Ge have been tested and different electronic properties were observed. The structure with minimum energy was selected for other calculations. The theoretical equilibrium total energy, lattice parameter $a_o$, and bulk modulus B are determined by fitting the total energy as a function of volume for all of the crystals to the Murnaghan's equation of state [6].

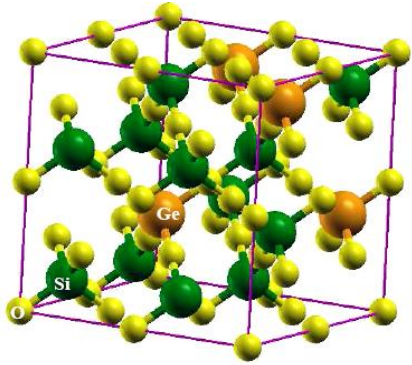

**Fig. 1**. $Si_{4-n}Ge_nO_8$ supercell for n=1.

Optical properties of a solid can be described by complex dielectric function $\varepsilon(\omega) = \varepsilon_1(\omega) + i\varepsilon_2(\omega)$. The imaginary part of the dielectric function $\varepsilon_2(\omega)$ is given by

$$\varepsilon_2(\omega) = \frac{Ve^2}{2\pi\hbar m^2 \omega^2} \int d^3k \sum_{ij} |<kn|\mathrm{p}|kn'>|^2 f(kn)(1-f(kn'))\delta(E_{kn} - E_{kn'} - \hbar\omega)$$

Where $\hbar\omega$ is the energy of the incident photon, **p** is the momentum operator, $|kn\rangle$ is a crystal wave function and $f(kn)$ is the Fermi distribution function. The real part of the dielectric function $\varepsilon_1(\omega)$ follows from the Kramer-Kronig relation [7].

## 3. Results and discussion

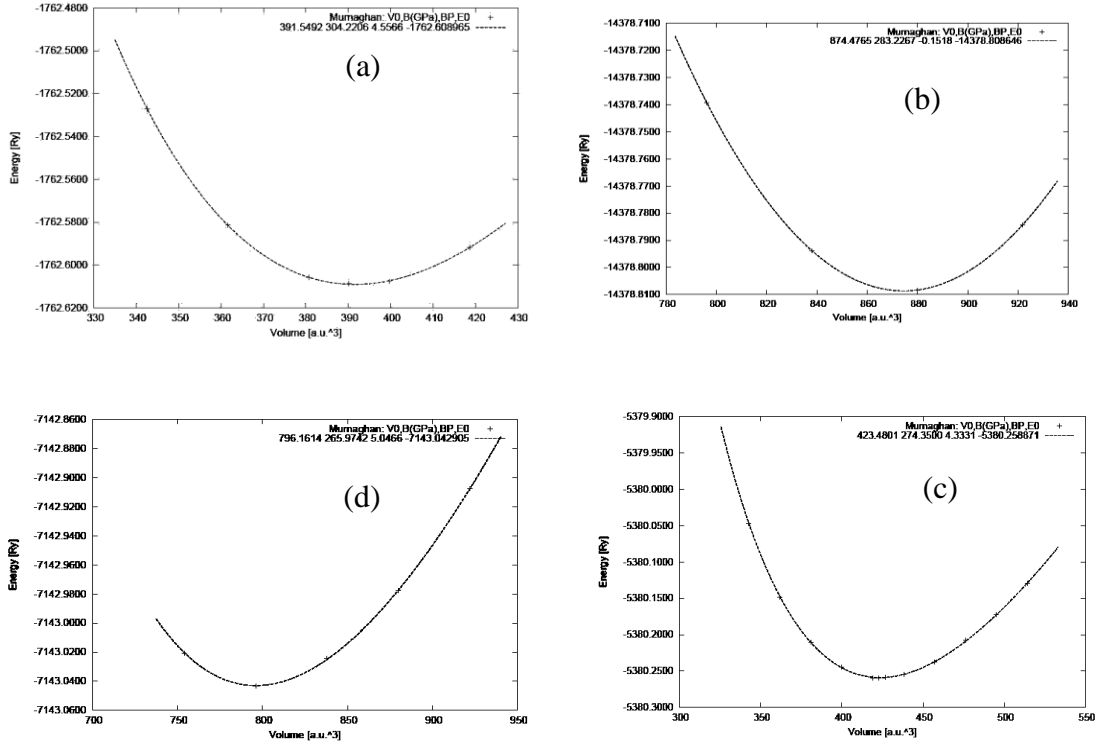

**Fig. 2**. Energy curves as a fonction of volume for (a) $SiO_2$, (b) $Si_{0.25}Ge_{0.75}O_2$, (c) $Si_{0.5}Ge_{0.5}O_2$, (d) $Si_{0.75}Ge_{0.25}O_2$.

The calculated energy curves as a function of volume for the all model compounds are shown in Fig. 2. The zero pressure equilibrium lattice constant ($a_o$), the bulk modulus ($B_o$) have been obtained for these compounds. Table 1 summarizes the results of our calculations compared with other theoretical results.

**Table 1.** Bulk modulus, equilibrium lattice constant, equilibrium and pressure band gap ($E_g$) for each crystal

|  | Bulk modulus B (GPa) | | | Equilibrium lattice constant (Å) | | |
| --- | --- | --- | --- | --- | --- | --- |
|  | This work (GGA) | Ref. [3] | | This work | Ref. [3] | |
|  |  | GGA | LDA |  | LDA | GGA |
| $SiO_2$ | 304.220 | 273 | 301 | 3.866 | 3.734 | 3.801 |
| $Si_{0.75}Ge_{0.25}O_2$ | 266.158 |  |  | 7.785 |  |  |
| $Si_{0.5}Ge_{0.5}O_2$ | 274.350 | 234 | 285 | 3.971 | 3.830 | 3.923 |
| $Si_{0.25}Ge_{0.75}O_2$ | 219.733 |  |  | 8.038 |  |  |

|  | $E_g$ (ev) on pressure | $E_g$ (ev) equilibrium | |  |
| --- | --- | --- | --- | --- |
|  |  | This work | Ref. [3] |  |
| $GeO_2$ | 3.972 |  | 2.402 | indirect |
| $Si_{0.25}Ge_{0.75}O_2$ | 4.160 | 2.395 |  | direct |
| $Si_{0.5}Ge_{0.5}O_2$ | 4.898 | 2.938 | 2.558 | indirect |
| $Si_{0.75}Ge_{0.25}O_2$ | 4.680 | 3.700 |  | direct |
| $SiO_2$ | 4.707 | 4.408 | 4.584 | indirect |

*3.1 Electronic properties*

The electronic properties of the $Si_{4-n}Ge_nO_8$ and $SiO_2$ are shown in Fig. 3. The results show that the indirect band gape of $SiO_2$ will become direct when Ge is added. The only exception is $Si_{0.5}Ge_{0.5}O_2$ which has an indirect band gape. As it is clear from the band structures, the degeneracy is removed after equilibrating the structures. Moreover, increasing the Ge concentration will decrease the nonequilibrium band gap except for the $Si_{0.5}Ge_{0.5}O_2$ which its band gap will increase by adding Ge to $Si_{0.75}Ge_{0.25}O_2$ (Fig. 4.) All of the equilibrium band gaps are smaller than their nonequilibrium values. As it is observed from Fig. 4 the equilibrium band gap will decrease by increasing the concentration Ge. The $Si_{0.25}Ge_{0.75}O_2$ and $GeO_2$ has the smallest band gap, but Bulutay et al. Show that $GeO_2$ is not stable [3]. $SiO_2$ has the largest band gap among compounds considered in this work.

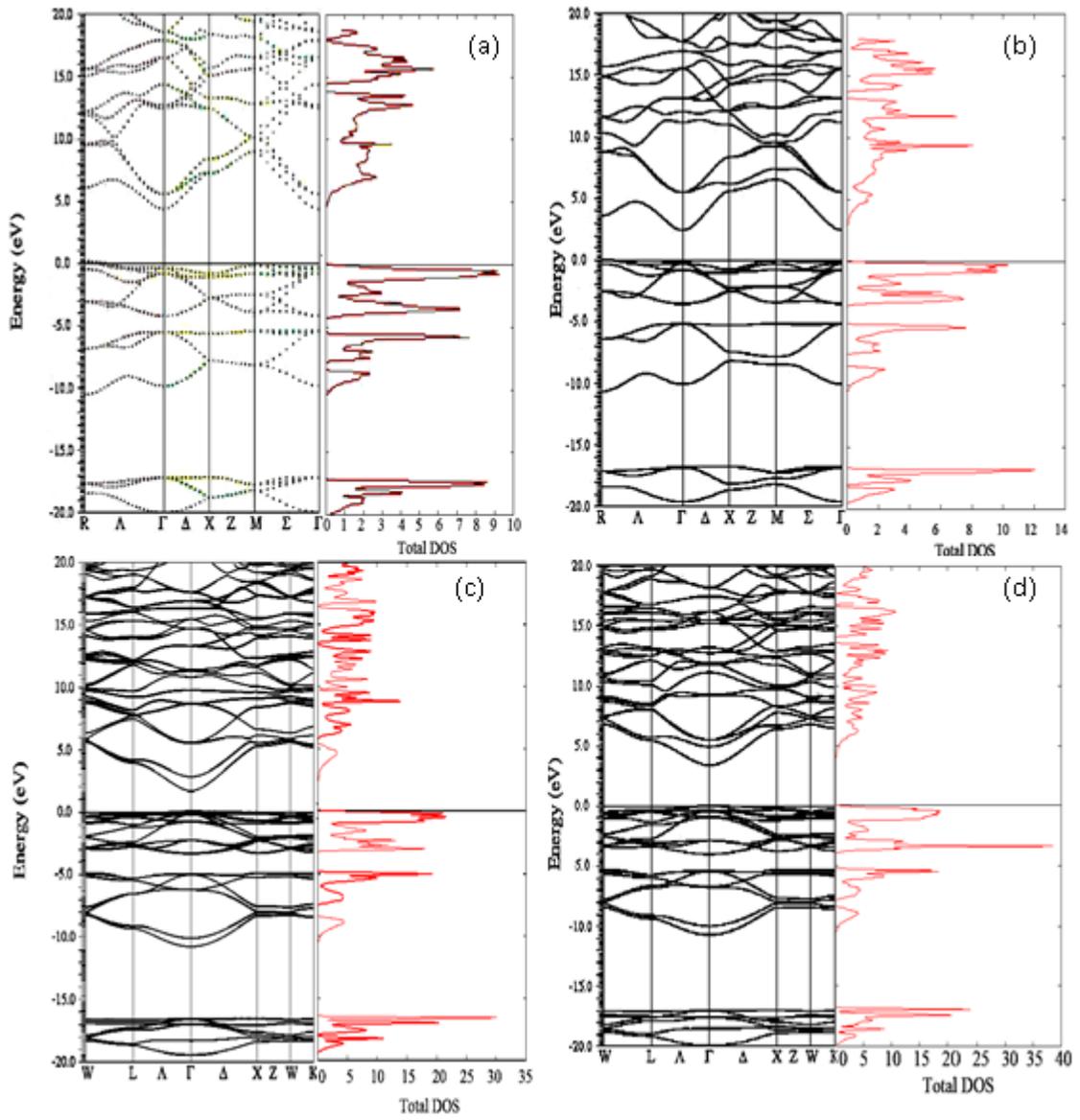

**Fig. 3**. Equilibrium band structure and DOS for (a) $SiO_2$, (b) $Si_{0.5}Ge_{0.5}O_2$, (c) $Si_{0.25}Ge_{0.75}O_2$, (d) $Si_{0.75}Ge_{0.25}O_2$.

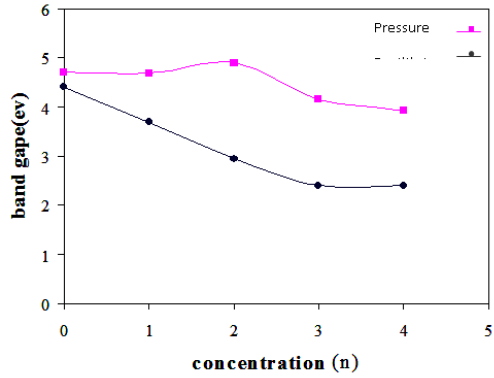

**Fig. 4**. Nonequilibrium and equilibrium band gap band gaps as a function of Ge concentration.

*3.2 Optical properties*

Optical calculation needs a dens k-mesh. 64 k-point seems to be sufficient for the supercells (n=1, 3). The scissors operator, which is the difference between the calculated and measured energy gape, was considered 1.0ev. This value was chosen approximately since there was not any experimental data. The imaginary and real parts of the dielectric function for all of the compounds are shown in Fig. 5.

Optical dielectric constant ($\varepsilon_1(0)$) which is $\varepsilon_1(\omega)$ at $\omega = 0$ was calculated with and without scissor operator (Table 2). As it is shown in Fig. 6, the calculations indicate that by increasing the Ge concentration, the optical dielectric constant is increased. This curve is not smooth and it increases suddenly from n=2 to n=3.

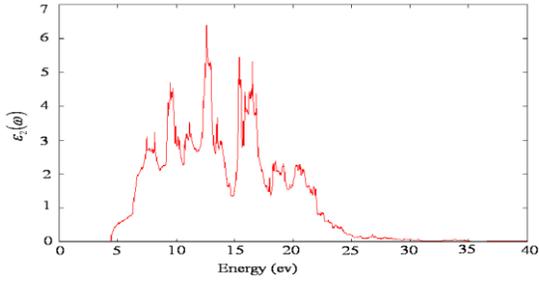
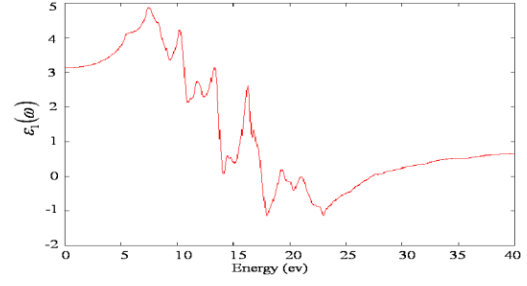

**(a)**

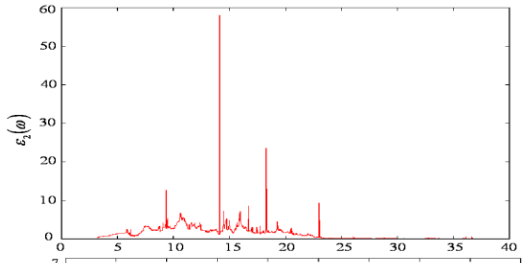
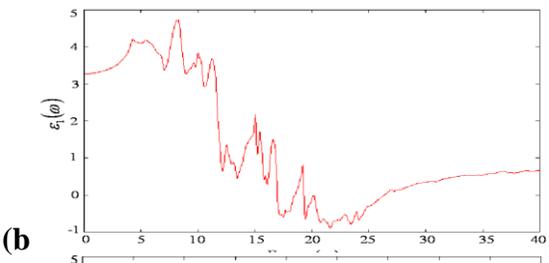

**(b)**

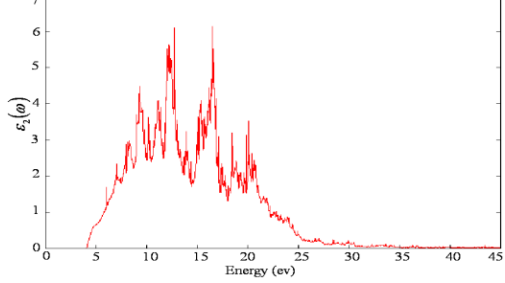
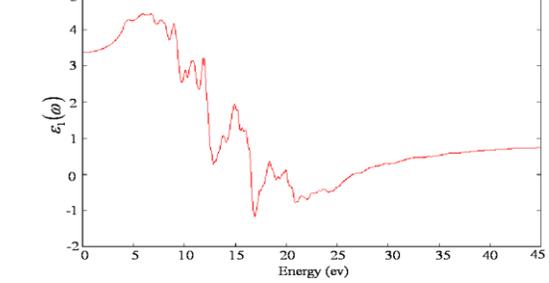

**(c)**

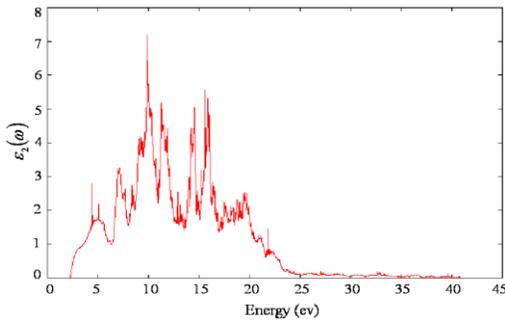
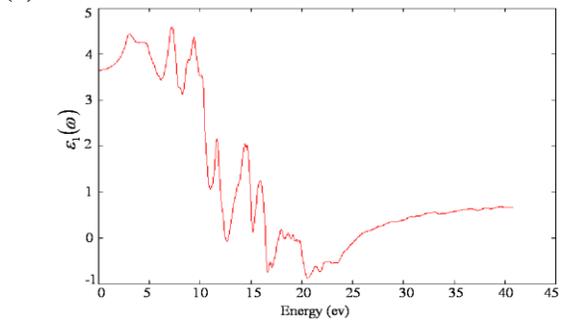

**(d)**

**Fig. 5**. Imaginary and real parts of the dielectric constant ($\varepsilon_1(\omega), \varepsilon_2(\omega)$) for (a) $SiO_2$, (b) $Si_{0.5}Ge_{0.5}O_2$, (c) $Si_{0.75}Ge_{0.25}O_2$, (d) $Si_{0.25}Ge_{0.75}O_2$

**Table 2.** Optical dielectric constant with and without the scissor operator

| $Si_{4-n}Ge_nO_8$ | n | $\varepsilon_1(0)$ with scissor operator | $\varepsilon_1(0)$ without scissor operator |
|---|---|---|---|
| $SiO_2$ | 0 | 3.21 | 3.31 |
| $Si_{0.75}Ge_{0.25}O_2$ | 1 | 3.263 | 3.37 |
| $Si_{0.5}Ge_{0.5}O_2$ | 2 | 3.274 | 3.520 |
| $Si_{0.25}Ge_{0.75}O_2$ | 3 | 3.820 | 3.820 |

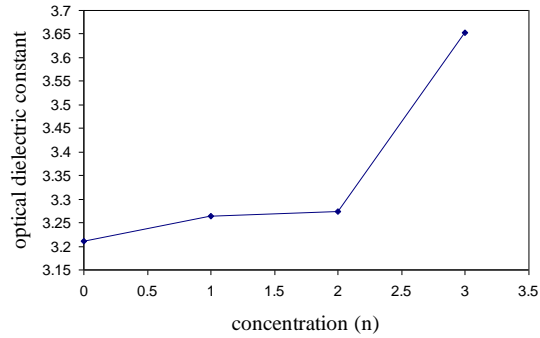

**Fig. 6**. Optical dielectric constant as a function of the Ge concentration.

## 4. Conclusions

To conclude, a more detailed first-principles calculation was carried out to study the electronic structure and optical properties of the $Si_{1-x}Ge_xO_2$ using the FPLAPW method. The ground state properties like equilibrium lattice constants, bulk modulus obtained. It was found that increasing Ge concentration and decreasing the Si concentration decrease the band gap and increase the optical dielectric constant. The $Si_{0.25}Ge_{0.75}O_2$ has the biggest optical dielectric constant and the smallest band gap. This band gap (about 2.39ev) is not small, so it can be used as a substrate for $Si_{1-x}Ge_x$ since it has a small band gap for some concentration of Ge [8].